\documentclass{ws-procs961x669}            
\usepackage{placeins}
\usepackage{float}
\usepackage{makecell}

\begin{document}
\title{High resolution calibration of string network evolution}

\author{J. R. C. C. C. Correia$^*$}
\address{Centro de Astrof\'{\i}sica da Universidade do Porto, and\\
Instituto de Astrof\'{\i}sica e Ci\^encias do Espa\c co, Universidade do Porto,\\ Rua das Estrelas, 4150-762 Porto, Portugal, and\\
Faculdade de Ciências da Universidade do Porto\\
 Rua do Campo Alegre 687, 4169-007 Porto, Portugal\\
$^*$E-mail: jose.correia@astro.up.pt}

\author{C. J. A. P. Martins$^+$}
\address{Centro de Astrof\'{\i}sica da Universidade do Porto, and\\
Instituto de Astrof\'{\i}sica e Ci\^encias do Espa\c co, Universidade do Porto,\\ Rua das Estrelas, 4150-762 Porto, Portugal\\
$^+$E-mail: Carlos.Martins@astro.up.pt}

\begin{abstract}
The canonical velocity-dependent one-scale (VOS) model for cosmic string evolution contains a number of free parameters which cannot be obtained ab initio. Therefore it must be calibrated using high resolution numerical simulations. We exploit our state of the art graphically accelerated implementation of the evolution of local Abelian-Higgs string networks to provide a statistically robust calibration of this model. In order to do so, we will make use of the largest set of high resolution simulations carried out to date, for a variety of cosmological expansion rates, and explore the impact of key numerical choices on model calibration, including the dynamic range, lattice spacing, and the choice of numerical estimators for the mean string velocity. This sensitivity exploration shows that certain numerical choices will indeed have consequences for observationally crucial parameters, such as the loop chopping parameter. To conclude, we will also briefly illustrate how our results impact observational constraints on cosmic strings.
\end{abstract}


\bodymatter

\section{Introduction}\label{aba:sec1}

Topological defects are expected consequences of phase transitions in the early Universe, formed via means of the Kibble mechanism \cite{Kibble:1976sj}. Depending on the symmetry broken different defects, with different dimension, can appear. The most studied defects are the filament-like cosmic strings. These are generically predicted in many candidates of Grand Unified Theories \cite{Jeannerot:2003qv} and superstring theory \cite{Sarangi:2002yt}, and are cosmologically safe, in the sense that they are not expected to overclose the Universe.

Due to their ubiquitous nature they have become a primary target for observational facilities, both current \cite{PlanckDefects,LIGODefects} and forthcoming \cite{CORE,LISA}. Analytical and observational studies of defect networks (strings included) often rely on a combination of semi-analytical modelling and simulations.

Here we present an up-to-date model calibration using high-resolution simulations, that ran in one of Europe's largest supercomputers, Piz Daint, using up to $4096$ GPUs. In order to ensure robustness, we will also explore how sensitive the model calibration is to certain numerical choices. We will also discuss our recently developed visualization strategy and avenues it might open in the study of defects. Some of this work has been published in Ref. \citenum{PhysRevD.10.2445}.

\section{Simulation Setup}\label{aba:sec2}

We begin with a brief description of our simulations. Take a Lagrangian density which, while originally invariant under $U(1)_L$ transformations, is already in the broken symmetry phase,
\begin{equation}
  \mathcal{L}=|D_\mu \phi|^2 - \frac{\lambda}{4}(|\phi|^2 -1)^2 - \frac{1}{4e^2}F^{\mu \nu}F_{\mu \nu}\,,
\end{equation}
where $\phi$ is a complex scalar field, $D_\mu \phi = \partial \phi - ieA_\mu$ are the covariant derivatives, $A_\mu$ is a gauge field, $F_{\mu \nu} = \partial_\mu A_\nu - \partial_\nu A_\mu$ is a gauge field strength and the scalar and gauge couplings are $\lambda$ and $e$, respectively. The defect network that will form corresponds to homotopically stable solutions of the equations of motion of this Lagrangian. In fact a cosmic a string is the result of a specific solution of the equations of motion, which maps $\phi$ to the symmetry breaking scale (here set to u; minima of potential) infinitely away from the string, and to zero at its core (maxima of potential). The equations of motion in Friedmann-Lemaitre-Robertson-Walker Universes and under the assumption of the temporal gauge ($A_0 = 0$) are
\begin{eqnarray}
\ddot{\phi} + 2\frac{\dot{a}}{a}\dot{\phi} &=& D^jD_j\phi -\frac{a^{2}\lambda}{2} (|\phi|^2 - 1)\phi \\
\dot{F}_{0j} &=& \partial_j F_{ij} -2a^2 e^2 Im[\phi^* D_j \phi]\,.
\end{eqnarray}
Abelian-Higgs field theory simulations of cosmic strings \cite{Bevis:2006mj,Hindmarsh:2017qff} merely use the discretization procedure of lattice gauge theory \cite{PhysRevD.10.2445}---which places scalar fields on lattice points distanced by $\Delta = 0.5$, and gauge fields on lattice links---to provide a discretized set of equations to update fields every conformal timestep. Due to the existence of Hubble damping and due to the topological nature of the potential, the initial conditions eventually relax into a configuration with multiple string networks. For simplicity, and to mimic what the fields are post-phase-transition, the initial conditions are set as follows: null on  every field, except $\phi$ which is set to have a random phase and unit magnitude.

We will additionally set scalar and gauge couplings to $\lambda = 2$ and $e=1$, respectively, which sets the Bogmolnyi ratio to $\frac{\lambda}{2e^2} = 1$ (a usual choice in the literature) and force the gauge and scalar core to have the same width. Given that the string radius varies as $r\propto \frac{1}{\sqrt{\lambda} a}$ and this could cause strings to "slip" through the lattice, we will make each coupling vary as
\begin{align}
    \lambda = \lambda_0 a^{2(1-\beta)} && e = e_0 e a^{(1-\beta)}
\end{align}
where $\beta$ is set to zero. This unphysical constant comoving width trick was first proposed in Ref. \citenum{PRS} and has been used throughout the literature to avoid resolution problems. We note that while previous work has suggested that $\beta=0$ and $\beta=1$ dynamics seem to agree, it is unknown presently if changing the value of $\beta$ would somehow pose a systematic error source in model calibration. This particular effect will be studied in a future publication and is beyond the scope of the present work.

In order to characterize the defect network that forms and its evolution throughout cosmic time, we also need to characterize the mean comoving distance between strings, $\xi$, and the mean squared velocity, $\langle v^2 \rangle$. For velocity estimators we will use two different forms, one first proposed in Ref. \citenum{Hindmarsh:2008dw} and the other in Ref. \citenum{Hindmarsh:2017qff}. The first one is based on a boosted static string (detailed derivation in Ref. \citenum{Hindmarsh:2017qff}) and will be referred to as the scalar field velocity estimator. It has the following form
\begin{align}
  <v^2>_{\phi} = \frac{2R}{1+R}, && R = \frac{\int |\Pi|^2  \mathcal{W} dx^3}{\int (\sum_i |D^+_{i} \phi|^2 ) \mathcal{W} dx^3}\, .
\end{align}
where $\mathcal{W}$ is weight function, chosen to localize the estimator on strings. We will use the Lagrangian density $\mathcal{L}$ to this end. Additionally, we will use the equation of state based velocity estimator,
\begin{equation}
  \label{eq:defEoS}
  <v^2>_{\omega} = \frac{1}{2} \bigg( 1+3\frac{\int p \mathcal{W} dx^3}{\int \rho  \mathcal{W} dx^3} \bigg)\,.
\end{equation}
We remark that the work of Reef. \citenum{Hindmarsh:2017qff} showed the second estimator to perform better in Minkowski space than the scalar based estimator, i.e. it is in better agreement with the analytical expectation for the velocity of an oscillating string in flat space.Such disagreement is lattice space dependent, and indeed an exploration of this feature in the opposite low velocity limit is published in Ref. \citenum{Correia:2021tok}.

One way to characterize the typical distance between strings is by counting the total amount of string in the box, $L$,
\begin{equation}
  \label{eq:defxiW}
  \xi_W = \sqrt{\frac{\mathcal{V}}{L}}\,,
\end{equation}
where $\mathcal{V}$ is box volume, and $L$ as mentioned the total amount of string. $L$ can be counted in multiple ways, but we will use an integer called winding. To explain it, remember one can think of a string as a magnetic flux tube, where the flux is quantized around the string,
\begin{equation}
    \Phi = \oint dx^\mu A_\mu = \frac{2\pi}{e}
\end{equation}
and the integer $n$ is the winding. If a cell face is pierced by a string, then the lattice  version of the winding \cite{Kajantie:1998bg} will be non-zero (1 for the choices of $\lambda$ and $e$ made above), and therefore we can say that a segment of length $\Delta x$ pierced the plaquette. Summing all the segments (taking care not to double-count) we then obtain the total length of string in the box, $L$.

Note that output of cells pierced by windings forms the basis of our visualization strategy. Given that this novel aspect of our simulation has thus far not been described elsewhere, we will now describe the strategy itself and what avenues it opens for the study of defect networks.

\subsection{Visualization strategy I - In-situ windings}

One of the most stringent and common bottlenecks for many High-Performance Computing simulations is related to Input/Output. Not only does computational throughput grow much faster than the typical read/write speeds of most storage solutions, but also, very often, the amount of data required to do science exceeds the amount of storage available. It is thus no surprise that in recent years there has been a push towards techniques of in-situ visualization, where the output data is heavily reduced beforehand and a smaller subset is used afterwards. There are many literature examples of successful application of such techniques \cite{CAMATA201823,8126857,10.1145/3332186.3332201,Sohrabi2019,AkiraKAGEYAMA2020}. In this section we will detail how this technique is applied in our multiGPU cosmic string simulations \cite{Correia:2020yqg} and the resulting performance gains.

In order to use an in-situ approach to outputting string positions in the lattice, we use ParaView Catalyst version 5.8.0. There are then two relevant components: the Adaptor code (written in C++) and a Python script which applies three filters to the data and then outputs an Unstructured grid (in *.pvtu format).

In the case of the Adaptor, it begins (if necessary) by creating a vtkMultiBlockDataset, where each rank contains then a vtkImageData with the correct information about block extent (in number of points), lattice spacing and origin of each block or sub-domain of the grid. Afterwards a series of vtkFloatArray are either created or updated by copying the contents of 7 different arrays pinned host memory buffers updated by GPUs before-hand. Six of these arrays (one for each cell face) contain only $\pm1$ in case a string pierces a cell face and $0$ otherwise. The last array is merely an $OR$ of the previous six, and will be used to determine which cells are pierced by strings and should therefore be outputted.

This determination is done in the Python script, which applies a Threshold filter to the data, selecting only cells wherein the last array is non-zero. In other words, from this point on we already have a reduced dataset. The Merge Block filter is then applied to collect pieces of strings spread throughout several sub-domain. To finalize the Parallel Unstructured Grid writer is used to output data in the *.pvtu format (alongside multiple block files *.vtu).

\begin{figure}
\begin{center}
\includegraphics[width=\columnwidth,keepaspectratio]{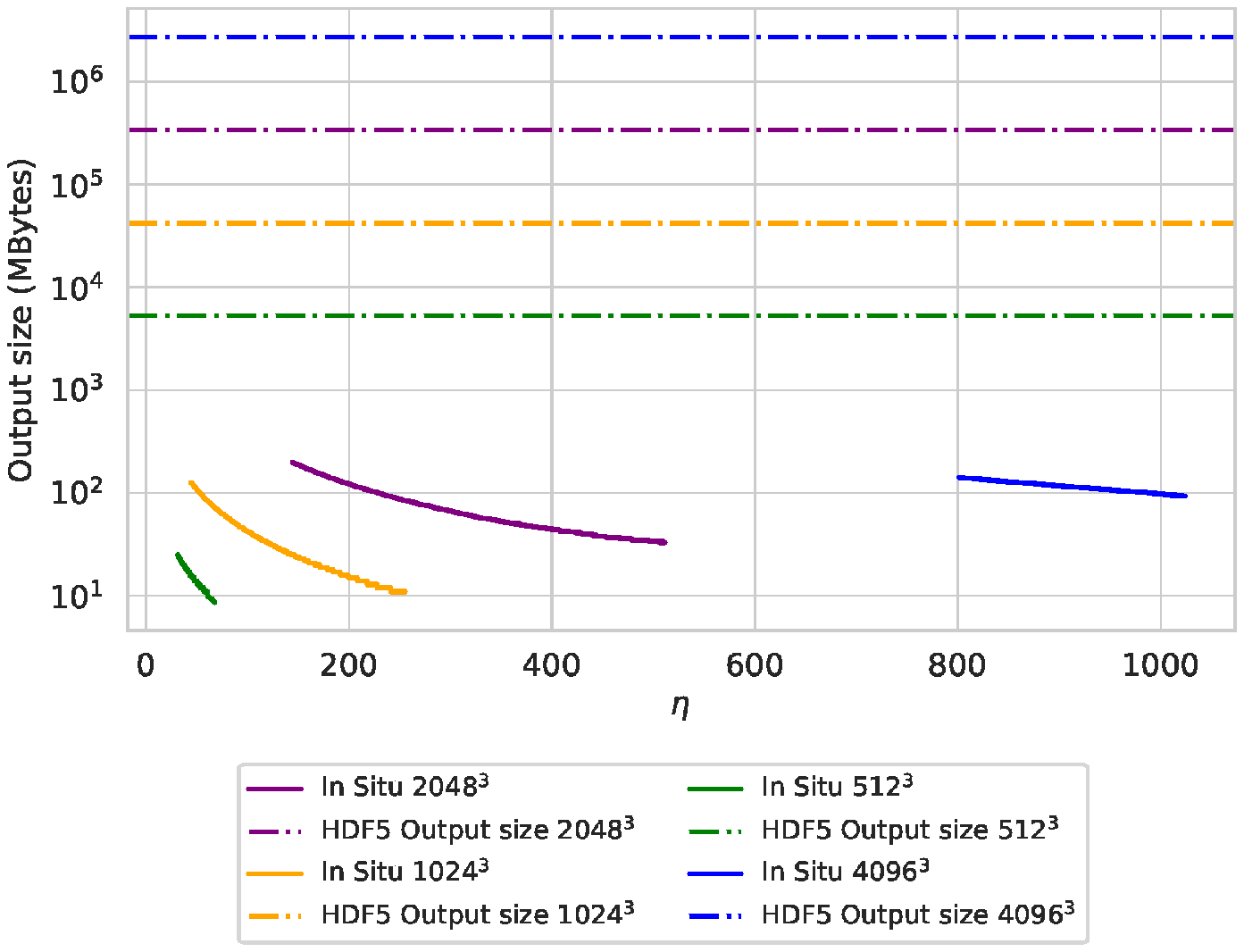}
\includegraphics[width=\columnwidth,keepaspectratio]{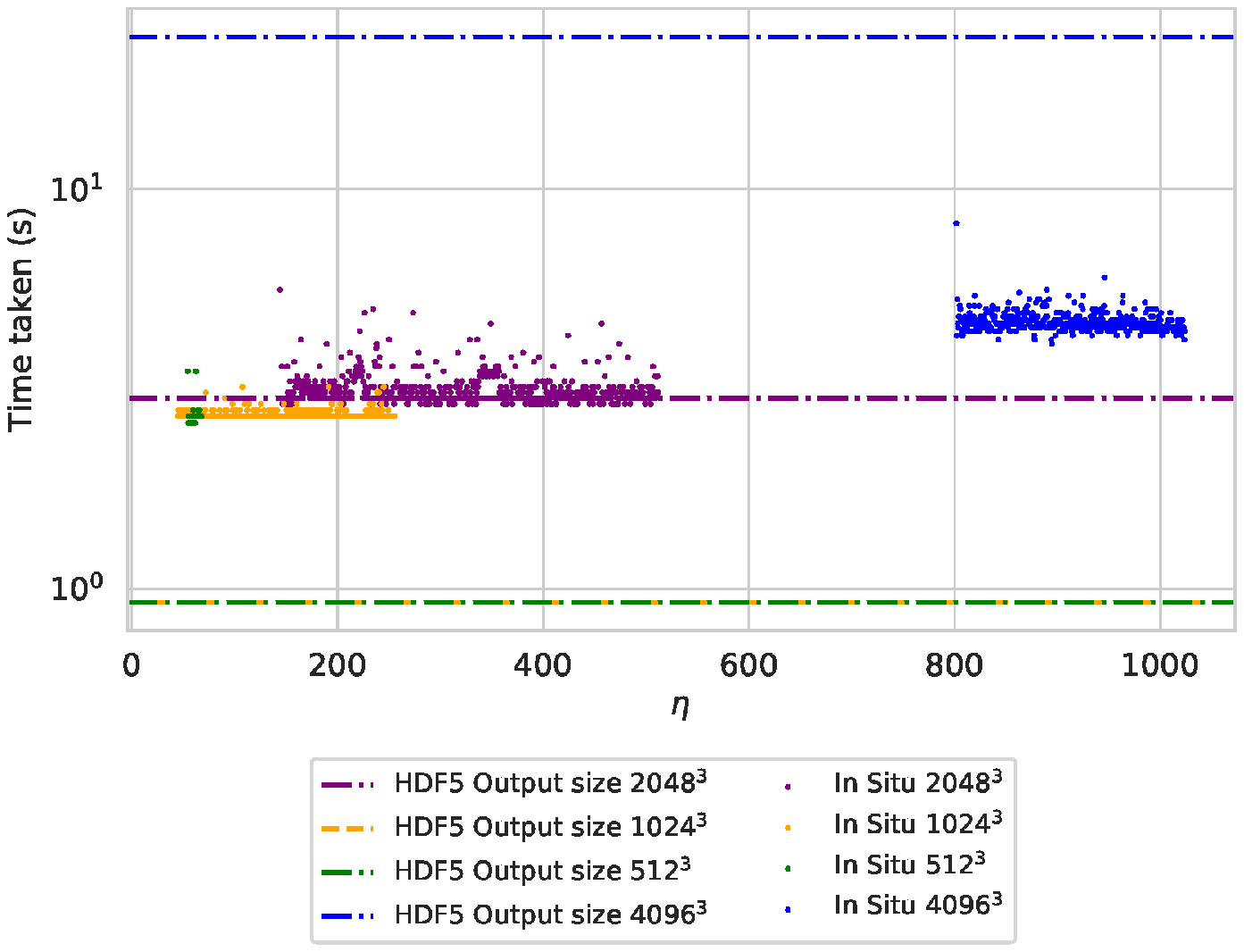}
\end{center}
\caption{Top: Size of raw HDF5 data output  (windings for six cell faces, dashed lines) and of that for the In-Situ approach, where Unstructured grids are output (full lines), for lattice sizes: $4096^3$ (blue), $2048^3$ (purple), $1024^3$ (orange) and $512^3$ (green). Bottom: corresponding output times.}
\label{iogains}
\end{figure}

\begin{table}
\tbl{A summary of typical output sizes for a timestep either with Raw output of all cells (HDF5) or using only the unstructured grid outputs from our in-situ approach.}
{\begin{tabular}{lcccc}
\toprule
  Lattice size &  \multicolumn{2}{c}{Output data size (MB)} & \multicolumn{2}{c}{Time taken (s)} \\
               &HDF5 estimated & In-situ measured & HDF5 estimated & In-situ measured \\
\colrule
  $512^3$  & 5296 & $[8.7, 25.0]$      & 0.9  & $[2.6, 3.5]$   \\
  $1024^3$ & 42368 & $[11.0, 126.0]$   & 0.9  & $[2.7, 3.2]$   \\
  $2048^3$ & 338944 & $[33.0, 199.0]$  & 3.0  & $[2.9, 5.6]$   \\
  $4096^3$ & 2711552 & $[99.0, 144.0]$ & 23.9 & $[4.1, 8.2]$ \\
\botrule
\end{tabular}}
\begin{tabnote}
Given that the in-situ approach size depends on the number of cells that are part of a string, we indicate the range of sizes obtained over the conformal time ranges where we output.
\end{tabnote}\label{tab:spacesaving}
\end{table}

Note that no rendering is done so far: the in-situ aspect outputs heavily reduced data, but it does not produce a string centerline. The next section will describe how to create such a centerline by connecting each cell of each individual string and applying an averaging to string positions, such that the effect of taxicab geometry is greatly reduced.

Before we move on, we must detail the gains obtained from using this particular approach to data output, by comparing to the previous method of output of the simulation, which was to output all cells throughout the domain in HDF5 files (file-per-process approach). All benchmarks were conducted at Piz Daint, the Swiss National Supercomputer, the $4th$ most powerful supercomputer in Europe at the time of writing \cite{TOP}. This machine features $5704$ nodes, each equipped with an Nvidia Tesla P100, together with a $scratch$ partition which is a Cray Sonexion 3000 Lustre filesystem with 8.8PB capacity. The system contains 40 object storage targets, and can handle a file per process approach well as long as there are not thousands of files in a single folder (this can be circumvented by grouping files according to rank for instance).

The first, more obvious consequence is how this results in less data being output. The smallest size of single HDF5 file with 7 single precision floating point datasets of dimension $(256, 256, 256)$ is around $662$ MB in our test set. In order to determine the total typical output size for the benchmarks we are about to describe, we need to multiply the number by the number of ranks used. Keeping the problem size per process fixed. We will then run the benchmarks for in-situ with the following lattice sizes $512^3$, $1024^3$, $2048^3$ and $4096^3$ with domain decomposition of $(2,2,2)$, $(4,4,4)$, $(8,8,8)$ and $(16,16,16)$. In all benchmarks we output data every five timesteps. The obtained total timestep output sizes are shown in Table \ref{tab:spacesaving} and in Figure \ref{iogains}. We note that as the networks evolve, and the strings interact, the energy density of strings decreases which has an immediate and obvious effect on the number of cells output: they decrease with conformal time. This is translated in the total timestep size, which decreases with conformal time for every lattice size. The other important conclusion is that there is a vast reduction in total storage space required, at worst two orders of magnitude, at best four. This not only allows us to output data at a larger temporal rate (more timesteps), but also allows the possibility of adding more data to the outputs, if necessary. 

We must also ask if there a performance (wall-clock time) speed-up or a performance penalty, or if, even including a costly filter like Merge Blocks, we can output data at similar rates as in the previous case. In order to do so, we will use the maximum measured writing bandwidth of our file per process approach with HDF5 files with 8, 64 and 512 writers for $512^3$, $1024^3$ and for both $2048^3$ and $4096^3$. The three maximum bandwidth are, respectively $5712$, $45640$ and $113300$ MB/s. Note that we will use the maximum measured bandwidth of 512 writers for 4096 as it is already very close to the estimated peak bandwidth of the Lustre filesystem on Piz Daint. Using these maximum bandwidths and the estimated storage needed for each file allows us to calculate a typical time taken. For in-situ we measure the time taken to output each timestep where data is written out. Our results, shown in table \ref{tab:spacesaving} and in figure \ref{iogains}, indicate that in-situ is definitely worth it if the file size becomes extremely large and the bandwidth is already saturated (close to the theoretical maximum bandwidth of the filesystem). This is evident when comparing the $2048^3$ and $4096^3$ cases where the amount of time taken is roughly comparable for $2048^3$ but in the $4096^3$ the amount of time taken for in-situ is roughly an order of magnitude below.

In conclusion, the technique described herein makes studying small-scale structure of strings possible with large lattices, not only reducing the amount of storage required by the simulation towards feasible amounts, but also permitting a significant speed-up, circumventing the limitation imposed by the maximum bandwidth of the filesystem.

\subsection{Visualization of string networks II - post-processing}

After the simulation runs, we can use the Unstructured grid output to construct string centerlines. The first step is to group neighboring cells via the Connectivity filter, and then use the PassArrays filter to pass all the data arrays to the input of the custom filter. The centerline script loops over each region identified by the connectivity filter and by each cell. The problem of trying to connect cells based on the order of the cellId in the Unstructured grid is that they are ordered by index (i,j,k) and not by the order they should be in a string.

The way to solve this issue is to begin in a cell in the string, ask for the neighbors according to the connectivity filter and then use the information contained in each of the winding arrays to understand if the connectivity established via cell faces is physically valid, ie. if magnetic flux passes through that cell face. Note that we choose the positive, $+1$ direction for the magnetic flux but the script would work if working in the opposite direction too. Once a valid neighbor is identified, the vortexBounds for the specific cell and the next one are used to find cell center coordinates. These coordinates used to trace a vtkLine connecting one cell to the next, passing in the middle of a cell face. Cell by cell and segment by segment, the string centerline begins taking shape, as a collection of vtkLines (hence the output is of vtkPolyDataCollection type). 

\begin{figure}
\begin{center}
\includegraphics[width=\columnwidth,keepaspectratio]{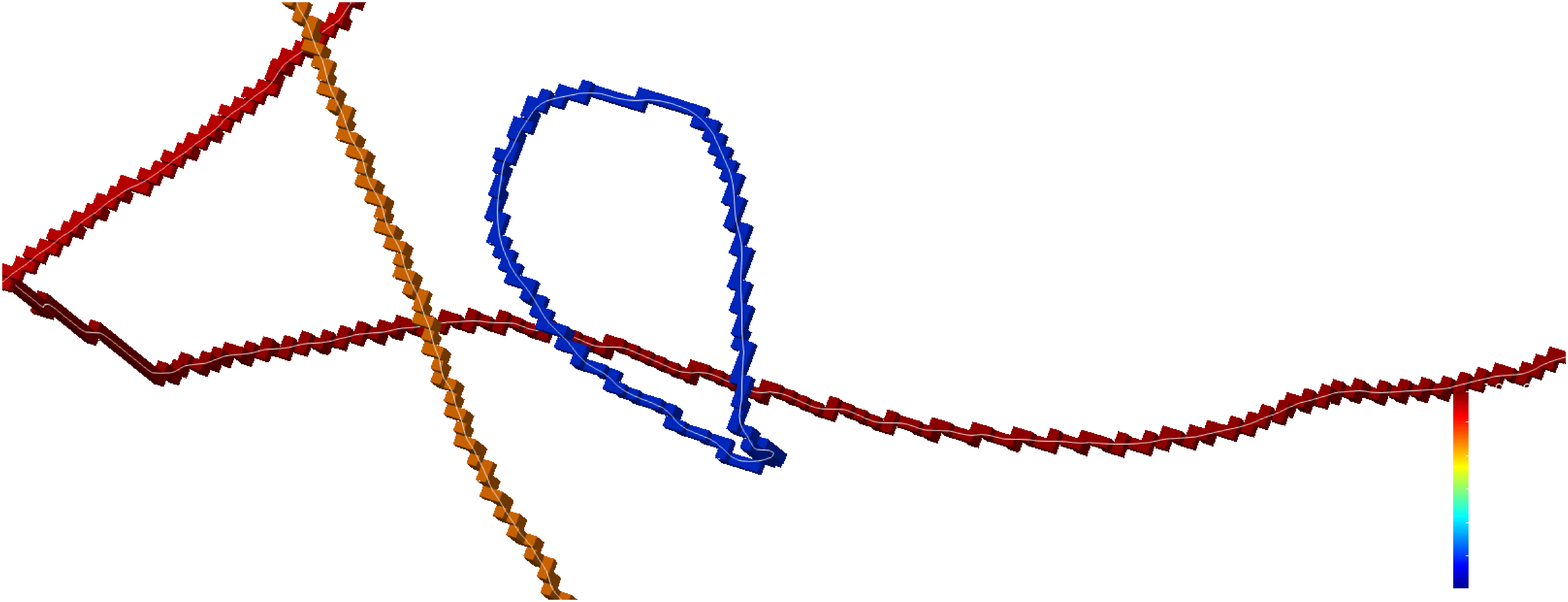}
\includegraphics[width=\columnwidth,keepaspectratio]{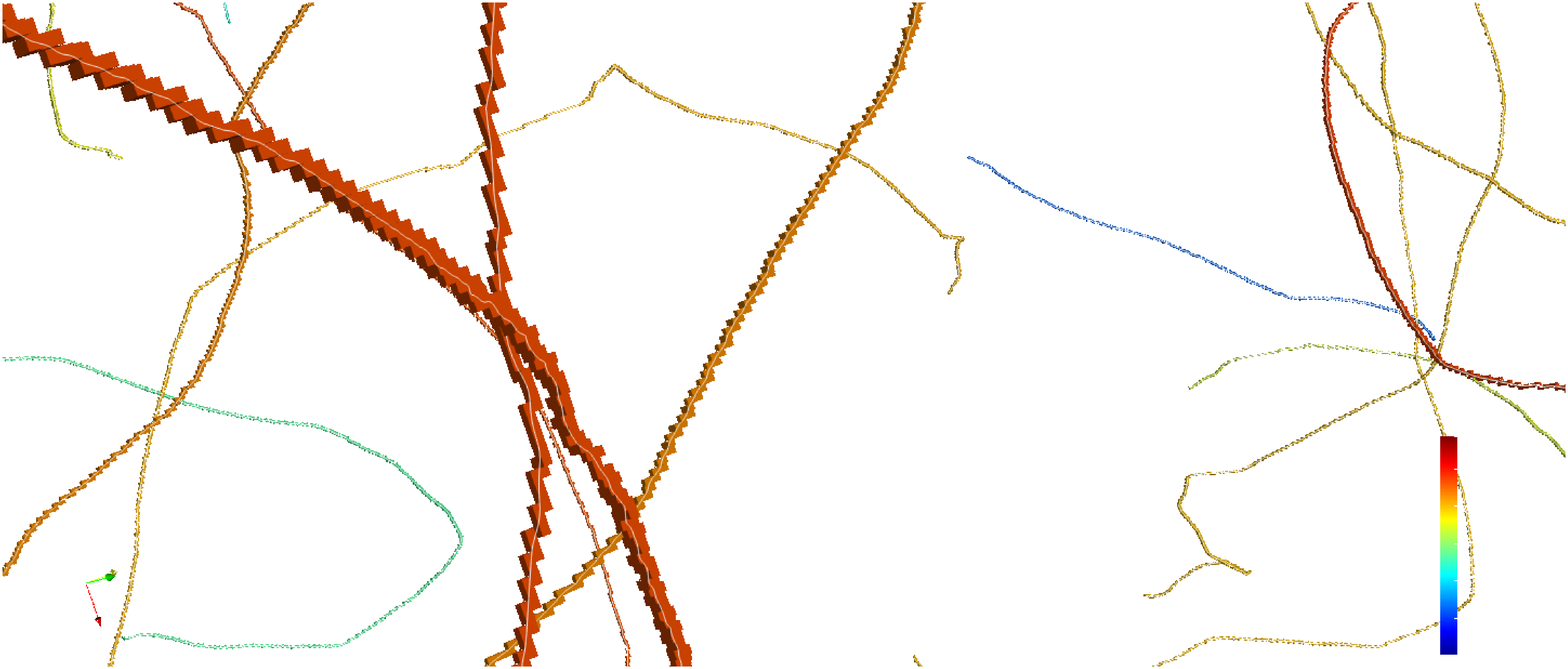}
\end{center}
\caption{String cells colored by region in two close-up screenshots of $2048^3$ radiation era simulation. We show in addition the output of the centerlines custom filter with smoothing via a Hanning window. A Loop (in blue) is shown at the center of the top panel screenshot. An intercomutation event (in red) is shown on the left-hand side of the bottom panel.}
\label{Loop}
\end{figure}

There are however some special cases to be mentioned as they can cause issues in the correct visualization and analysis.These are merely a consequence of the phenomenology of a string network. When two strings meet at one point (forming an X-shaped region), they can exchange ends, i.e. intercommute, and two new strings emerge immediately afterwards. On the other hand, if strings are to either meet in two points or to self-intersect, it is very likely that a closed loop forms. Loops are extremely important for observational consequences, as they shrink and dissipate energy\footnote{Here lies a disagreement between two types of cosmic string simulation in the literature: in what type of energy do loops decay? Gravitational waves, massive radiation? This can alter the overall energy loss of the network, subsequent evolution of the network, and observational imprints. The scientific goal of the numerical techniques developed in the present work is to address this issue.} therefore generating some observational fingerprints.

Intercommutation events (or X-regions) can cause the string centerline to stop at the boundary of the box but not go through all the cells of the entire region. In effect the initial version of the custom filter would draw a single string (corresponding to two legs of the X-region, so to say). This can be avoided by comparing the number of cells used for the creation of the centerline so far and the total number of cells in the region. If one is much smaller, then the centerline drawing must restart at a cell not used previously and create the other two legs. Loops require us to close the path of the centerline, in other words, to connect the last cell and the first. Identifying whether a loop is present can be done using the number of points and the number of cells. Should they differ by one, it is necessary to close this path.

\begin{figure}
\begin{center}
  \includegraphics[width=\textwidth]{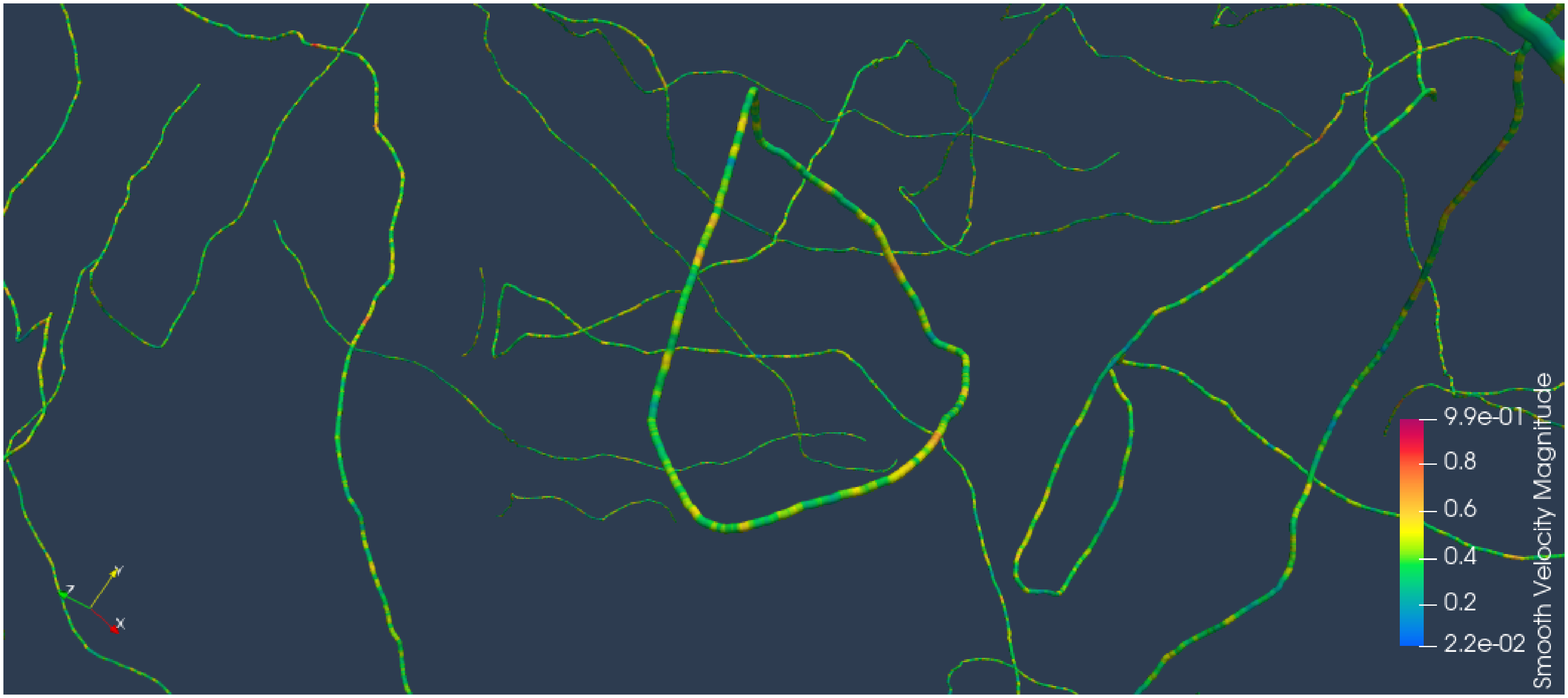}
  \includegraphics[width=\textwidth]{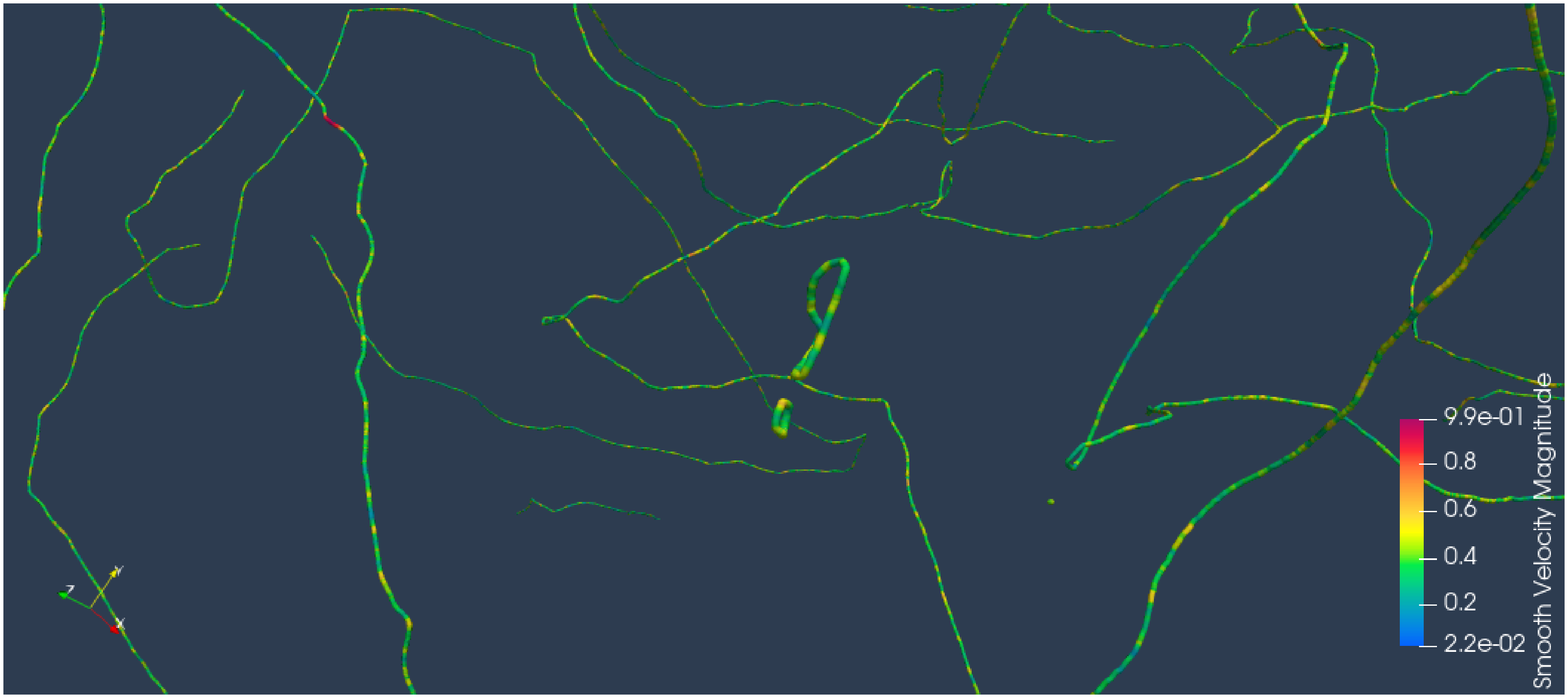}
  \includegraphics[width=\textwidth]{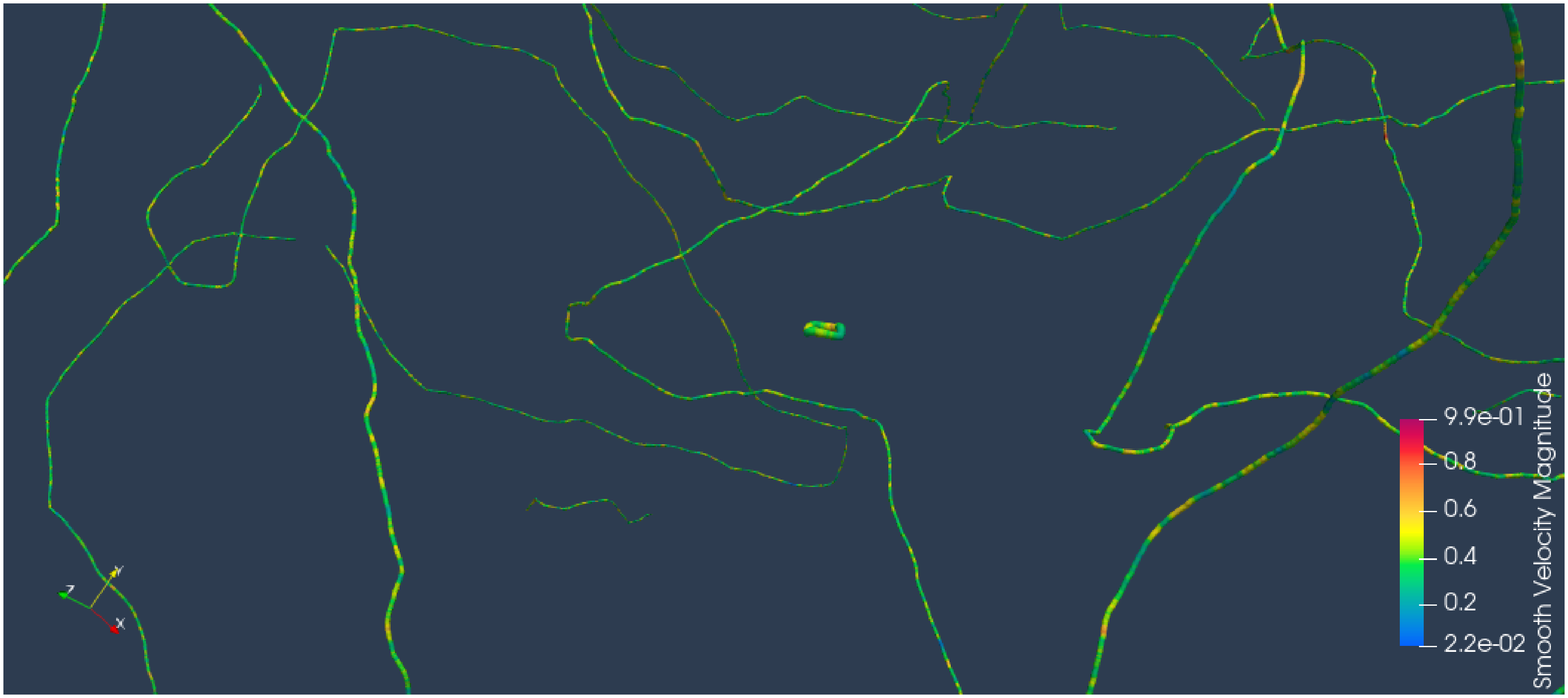}
\end{center}
\caption{Comparison of the mean rate of change of correlation length $\xi / \eta$ (top left) and the mean velocity $\langle v \rangle$ (top right) with the solid lines corresponding to the calibration and the shaded regions to the uncertainty of the measurements of each estimator for three different box sizes. The bottom plots show how these differences impact the momentum parameter $k(v)$ (bottom left) and in the energy loss parameter $F(v)$ (bottom right).}
\label{figCenterlines}
\end{figure}

After dealing with these two problematic cases we now have the string positions along every stair-case-like segment.This staircase behavior is no more, no less a consequence of the discrete nature of the lattice and of not having complete information about where exactly in each cell face a string lies (one assumes perfectly in the middle but this is not entirely realistic either). There is therefore some small-scale structure on the resulting strings in a scale roughly of the lattice size. Given that this is not a physical effect, we wish to remove it. In Ref. \citenum{Hindmarsh:2008dw} the smoothed string was obtained by averaging string positions along the string path. Here we convolve each string with a window function in order to remove such artificial structure. The specific window function and the window size can be chosen in the filter. Given that numpy and scipy are used for this procedure, possible window functions by default include Hanning, Haming, Bartlett, Blackman. Figure \ref{Loop} shows a collection of cells in two long strings and one loop either in cells as the resulting output of the Connectivity filter or as string centerlines (in white) smoothed over with the Hanning window function. Qualitatively this is more representative of a natural string. An example of the completed centerlines for a $4096^3$ lattice simulation, with $\Delta x=0.5$ aand $\Delta \eta=0.1$ in radiation epoch is shown in Fig. \ref{figCenterlines}.

\section{Extended Velocity-dependent One-Scale model}\label{aba:sec3}

The canonical semi-analytic model for string network evolution is the so-called Velocity dependent One-Scale model of Refs. \citenum{Martins:1996jp,Martins:2000cs}. It can be thought of as thermodynamic model in the sense that it describes the time evolution of two network averaged quantities, the mean string separation $L$, and the root mean squared velocity $v$. The cosmological history of these quantities is described in terms of relevant physical effects. For instance, the expansion of the Universe tends to act as a friction force on the strings (thereby affecting velocities) and also acts to dilute the network (affecting the mean separation). The model takes the following form, in comoving coordinates,
\begin{eqnarray}
\frac{d\xi}{d\eta} &=& \frac{m\xi}{(1-m)\eta} v^2 +F(v)  \label{eq:vos1_comoving}\\
\frac{dv}{d\eta} &=& (1-v^2) \bigg[ \frac{k(v)}{\xi}  - \frac{2m v}{(1-m)\eta} \bigg],.\label{eq:vos2_comoving}
\end{eqnarray}
where $\xi=La$ is the comoving mean string separation, $\eta$ is the conformal time, related to $t$ by means of $d\eta= adt$, $\mathcal{H}$ is the conformal Hubble parameter, $m$ is an expansion rate such that the scale factor $a\propto t^{m}$, and $k(v)$ and $F(v)$ are two phenomenological functions, related to the curvature of strings and to energy loss processes, respectively. In the original VOS, $k(v)$ can be analytically determined by comparison with the helicoidal string ansatz. However, for the present work, we will allow a more general form,
\begin{equation}
  k(v) =  k_0\frac{1-(qv^2)^{\beta}}{1+(qv^2)^{\beta}}.
\end{equation}
where $q$, $k_0$ and$\beta$ are free parameters, which in the case of $q$ and $k_0$ have a clear physical meaning. The former corresponds to the inverse of the maximum defect velocity squared $q \approx 1/v^2$ and the latter to the maximal value of $k(v)$, obtained in the low velocity limit. Note that the helicoidal ansatz corresponds to setting $q=2$, $k_0 = \frac{\sqrt{2}\pi}{2}$ and $\beta=3$. Letting these parameters free allows us to verify if small-scale structure on strings deviates from the Nambu-Goto expectation significantly, as this would be reflected on parameters $q$, $k_0$ and $\beta$ taking different values from the analytical expectation.

The energy loss function $F(v)$ was originally devised to only include a term linearly proportional to the velocity to model energy loss via loop chopping. However, recent work on radiation on Abelian-Higgs strings \cite{Hindmarsh:2017qff}, then prompted the authors of Ref. \citenum{Rybak1} to introduce a radiative loss term when studying domain wall evolution. Such a term is described by a power law of curvature,
\begin{equation}
  F(v) = cv +  d[k_0-k]^r\,.
\end{equation}
where $c$, $d$ and $r$ are free parameters which express the loop chopping parameter and the energy loss normalization and exponent, respectively. Although the original VOS, with the helicoidal ansatz and no explicit energy loss, had only one parameter, the extended version has 6. This is not a problem per se, as long as one can extract the correct velocity dependencies of each function $k(v)$ and $F(v)$. It is for this reason we will simulate string networks in Universes where with a power law scale factor, $a \propto m$ and with expansion rates $m$ in the range $m \in [0.5,0.95]$ (i.e. not only matter and radiation epoch are simulated).

We note that since string networks, under any Universe with a power law scale factor ($a\propto t^m \propto \eta^{m/(1-m)}$), are expected to evolve according to the so-called linear scaling regime where,
\begin{align}
  \xi \propto \eta  \propto d_H && v = const \, ,
\end{align}
we can easily use the measured $\dot{\xi} \sim \frac{\xi}{\eta}$ and $v$ to extract the proper velocity dependencies of $k(v)$ and $F(v)$, and also calibrate model parameters. To this end, we will use the same Markov Chain Monte Carlo pipeline of Ref. \citenum{Correia:2020gkj} to assess posterior distributions on all model parameters, and thus unveil possible correlations, and predict likelihood maxima and uncertainties. All model parameters are sampled from uniform distributions, with the range of these distributions adapted to each dataset as necessary. The logarithm of the likelihood is to be given by the $\chi^2$ statistic. We will use 32 walkers and a minimum of $10000$ steps.

\section{Dynamic range and lattice size}\label{aba:sec4}

String simulations have a problem of separation of scales, as it is often necessary to resolve scales all the way from the size of the horizon, down to scales comparable to the radius of the strings. When it comes to better resolving small-scale structure, there are two possible ways to do so, either by decreasing lattice spacing or by increasing lattice size, which in turn increases dynamic range. The latter follows because the lattice size directly determines the final simulation time: half-a-light crossing time or when the horizon size is half the box size. If a larger horizon can be reached by the end of the simulation, then for the same lattice spacing we can better resolve scales of a fraction of the horizon.

\begin{figure}
\begin{center}
  \includegraphics[width=\textwidth]{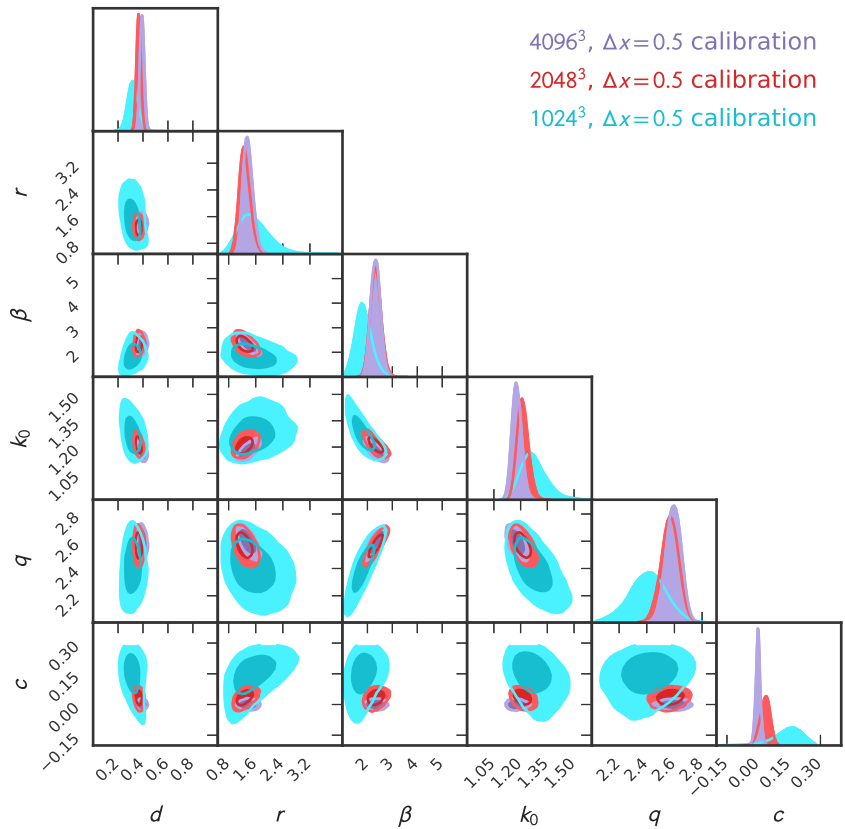}
\end{center}
\caption{Corner plots for the MCMC calibration of the VOS model, obtained with the velocity estimator $\langle v_\omega \rangle$, lattice spacing $\Delta x = 0.5$, box sizes $4096^3$, $2048^3$ and $1024^3$. Remake of figure 3 of Ref. \citenum{Correia:2021tok}.}
\label{figCorner1}
\end{figure}

We now explore the effects of increasing lattice size, while keeping spacing constant throughout ($\Delta x = 0.5$). To this end, we calibrate the VOS with $1024^3$, $2048^3$ and $4096^3$ lattices, with expansion rates $m\in[0.45, 0.95]$, 10 simulation runs each. For now we will use the equation of state velocity estimator. The resulting calibrations are summarized in table \ref{table1} and the corresponding posterior plots are found in figure 3 of Ref. \citenum{Correia:2021tok}, remade in fig. \ref{figCorner1}.

\begin{table}
\tiny
\tbl{Calibrated VOS model parameters for our three different lattice sizes, $1024^3$, $2048^3$ and $4096^3$, all with the same lattice spacing $\Delta x=0.5$, and two different choices of velocity estimators, $\langle v^2_\omega \rangle$ and $\langle v^2_\phi \rangle$ (in the top and bottom parts of the table, respectively),}
{\begin{tabular}{lcccccccc}
\toprule
\thead{Lattice\\size} & $\Delta x$ & \thead{Velocity\\estimator} & d & r & $\beta$ & $k_0$ & q & c \\
\colrule
$1024^3$ &     &                              & $0.32^{+0.04}_{-0.04}$ & $1.51^{+0.48}_{-0.37}$ & $1.82^{+0.34}_{-0.30}$ & $1.27^{+0.08}_{-0.06}$ & $2.41^{+0.13}_{-0.13}$ & $0.15^{+0.05}_{-0.07}$ \\
$2048^3$ & 0.5 & $\langle v_\omega^2 \rangle$ & $0.37^{+0.02}_{-0.02}$ & $1.27^{+0.17}_{-0.15}$ & $2.33^{+0.21}_{-0.20}$ & $1.21^{+0.03}_{-0.03}$ & $2.57^{+0.06}_{-0.06}$ & $0.03^{+0.02}_{-0.03}$  \\
$4096^3$ &     &                              & $0.39^{+0.02}_{-0.02}$ & $1.36^{+0.15}_{-0.13}$ & $2.32^{+0.20}_{-0.18}$ & $1.18^{+0.03}_{-0.03}$ & $2.59^{+0.05}_{-0.05}$ & $0.00^{+0.01}_{-0.01}$  \\
\colrule
$1024^3$ &     &                              & $0.35^{+0.23}_{-0.10}$ & $2.39^{+1.58}_{-0.94}$ & $2.79^{+0.73}_{-0.56}$ & $1.06^{+0.05}_{-0.05}$ & $2.95^{+0.18}_{-0.19}$ & $0.44^{+0.04}_{-0.05}$ \\
$2048^3$ & 0.5 & $\langle v_\phi^2 \rangle$   & $0.33^{+0.05}_{-0.04}$ & $1.86^{+0.39}_{-0.32}$ & $2.65^{+0.28}_{-0.26}$ & $1.05^{+0.03}_{-0.03}$ & $2.84^{+0.08}_{-0.08}$ & $0.31^{+0.02}_{-0.02}$  \\
$4096^3$ &     &                              & $0.36^{+0.03}_{-0.03}$ & $1.72^{+0.26}_{-0.23}$ & $2.50^{+0.21}_{-0.20}$ & $1.06^{+0.02}_{-0.02}$ & $2.83^{+0.06}_{-0.06}$ & $0.23^{+0.01}_{-0.01}$  \\
\botrule
\end{tabular}}
\begin{tabnote}
Displayed values correspond to 16th, 50th, 84th percentiles of the posterior distributions.
\end{tabnote}\label{table1}
\end{table}

We see that most parameters remain relatively unchanged by increasing the lattice size, being subject only to a reduction of uncertainties (smaller contours in Fig. \ref{figCorner1}). The only exceptions are parameters $c$ and (to a lesser extent) $d$. We remark as well that these differences are qualitatively larger when comparing $1024^3$ with $2048^3$, than if comparing $2048^3$ to $4096^3$, which might hint that there is a minimum lattice size/dynamic range for model calibration. Gradually, $c$ is even reduced to zero at $4096^3$ which seems to suggest that as lattice resolution increases, loop-chopping is eventually replaced by radiative losses, bu this is at odds with information from visually seeing the network evolve---as loops are formed at various instants and at various sizes---as can be inferred from Fig. \ref{figCenterlines}.

In Ref. \cite{Correia:2021tok} we found, by comparing asymptotic quantities at all expansion rates, that the main culprit for the changes we discussed in the previous paragraph was the behavior of $\xi/\eta$. In fact, while the velocities remain mostly unchanged, $\xi/\eta$ is decreases (this is present in the literature too \cite{Bevis:2006mj,Hindmarsh:2008dw}) with lattice size, as does it uncertainty. The reduction of the uncertainties is responsible for the narrower posterior distributions, and the lower mean value translates itself in change in a downwards shift of $F(v)$. This is reflected in the parameters that control the normalization of $F(v)$, ie. in $c$ and (to a lesser extent), via anti-correlation $d$. Again the differences are larger when going from $1024^3$ to $2048^3$ than from $2048^3$ to $4096^3$.

Going back to our previous discussion of $c$, it appears the calibration suggested the loop-chopping parameter would eventually go to zero, however the visualization strategy showed loops being produced at different sizes and different size. Perhaps, as a cross-check, we should also repeat this analysis using the scalar field velocity estimator. The resulting parameters are summarized in the bottom half of table \ref{table1} and in figure 4 of Ref. \citenum{Correia:2021tok}. This now leads to a surprisingly different calibration, with very different model parameters, namely $c$ is statistically non-zero. This prompts an investigation on the biases of each velocity estimator and on possible solutions.

\section{Lattice spacing and velocity estimators}\label{aba:sec5}

Here we investigate why both velocity estimators give vastly different calibrations and suggest one possible solution. From the point-of-view of the extended VOS, the high expansion rates can be some of the most important (statistically) for two reasons. First, in the low-velocity limit, both velocity functions are reduced to one parameter each,
\begin{align}
    \lim_{v\to 0} F(v) = cv && \lim_{v\to 0} k(v) = k_0\, ,
\end{align}
and this means that overall the model becomes reduced to only two parameters. Uncertainties on observed asymptotic quantities also decrease with decreasing velocity (increasing expansion rate). Statistically this will mean that the high expansion rate limit will be important in determining $c$ and $k_0$, which will in turn affect all variables correlated with them, and that the lower uncertainties give a larger statistical weight to higher expansion rates. This leads us to suspect that the high expansion rate behavior of both velocity estimators might be the cause, as the difference between them increases with expansion rate \cite{Hindmarsh:2017qff}.

If we compare velocity estimators in the range $m\in[0.5, 0.95]$ (see figure 3 of Ref. \citenum{Correia:2019bdl}) the difference is maximal at $m=0.95$, and of about $\sim 10\%$. For even larger expansion rates, in the range $m \in [0.95, 0.997]$ (left panel of figure 5 of Ref. \citenum{Correia:2021tok}), the velocity estimators can disagree up to $60\%$ in the worst case scenario for lattice spacing $\Delta x=0.5$, In the work of Ref. \citenum{Hindmarsh:2017qff}, which explored the differences between the two estimators in Minkowski space (the opposite limit), it was shown that lattice spacing could reduce this disagreement. But even at high expansion rate this remains true: the disagreement is heavily reduced when we halve lattice spacing $\Delta x =0.25$ as can be seen on the right panel of the aforementioned figure 5. There is however one interesting difference between two opposite limits of high expansion rate and Minkowski: in the first the equation of state estimator approximates the scalar field, in the latter the opposite occurs. There is no a priori reason as to why both should behave the same way under lattice spacing reductions, therefore this is not in contradiction with previous results.

\begin{figure}
\begin{center}
  \includegraphics[width=\textwidth]{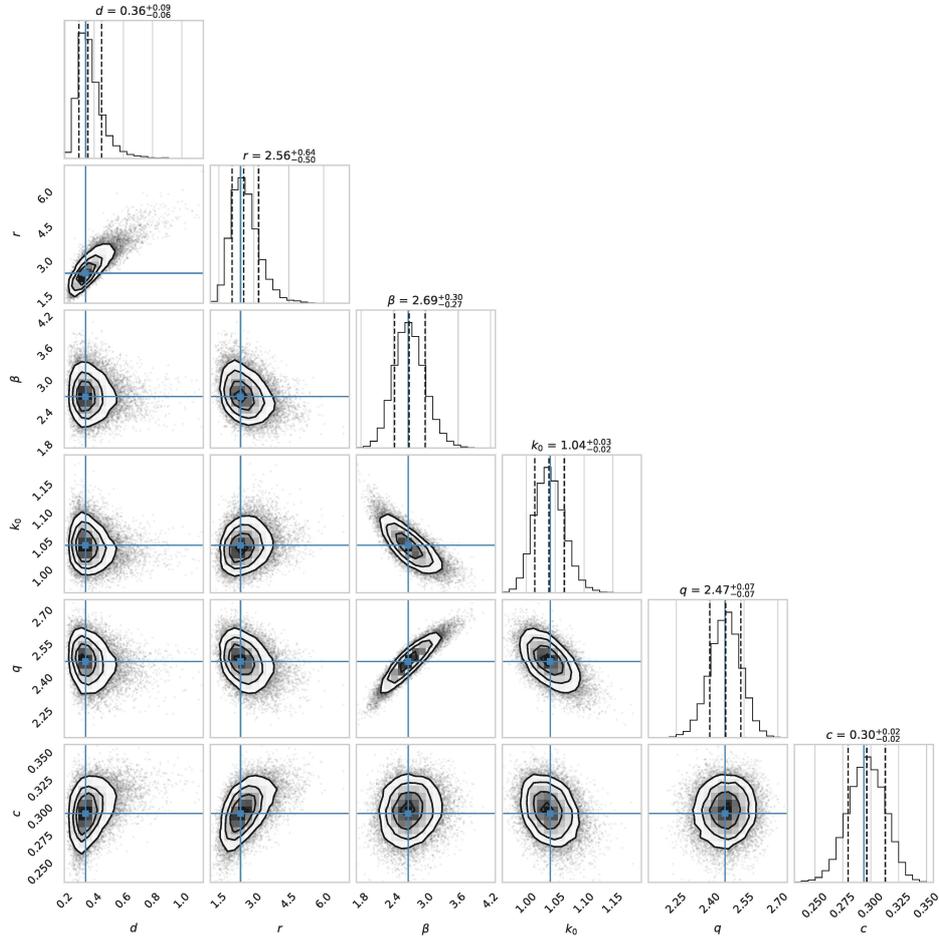}
\end{center}
\caption{Corner plots for the MCMC calibration of the VOS model, obtained with the velocity estimator $\langle v_\phi \rangle$, lattice spacing $\Delta x = 0.25$, box size $4096^3$.}
\label{figCorner2}
\end{figure}

As expected, the differences and their reduction under lattice spacing, are also reflected in the velocity functions $k(v)$ and $F(v)$. This as can be seen in figure 6 of Ref. \citenum{Correia:2021tok}, where two conclusions can be drawn. First, the equation of state velocity estimator yields unphysical $F(v)$ at lattice spacing $\Delta x = 0.5$, leading one to question its validity at the low velocity limit. Second, as the lattice is made finer, this brings both estimators in better agreement, namely by approximating $F(v)$ and $k(v)$ obtained via the equation of state estimator to the scalar field estimator. Note that this doesn't solve the problem of unphysical $F(v)$ completely, although for our intents and purposes it brings agreement for expansion rates $m\in[0.93,0.95]$. 

Now we compare the resulting model parameters for the two choices of lattice spacing. The results are summarized in table \ref{table2} and in the corresponding posterior contours of figures figure 7 and 8 of Ref. \citenum{Correia:2021tok}. We also showcase the contours of the $4096^3$, $\Delta x = 0.25$, $\langle v^2_\phi \rangle$ calibration in Fig. \ref{figCorner2}. Since the equation of state velocity estimator is highly sensitive to lattice spacing at large expansion rate, it is no surprise that the calibration drastically changes for all model parameters under a reduction of spacing, as can be seen both in the table and in the first aforementioned figure of Ref. \cite{Correia:2021tok}. With the finer calibration, it is also obvious the value of $c$ is no longer consistent with zero, that indeed loop-chopping is still a viable energy loss mechanism.

\begin{figure}
\begin{center}
  \includegraphics[width=0.49\columnwidth]{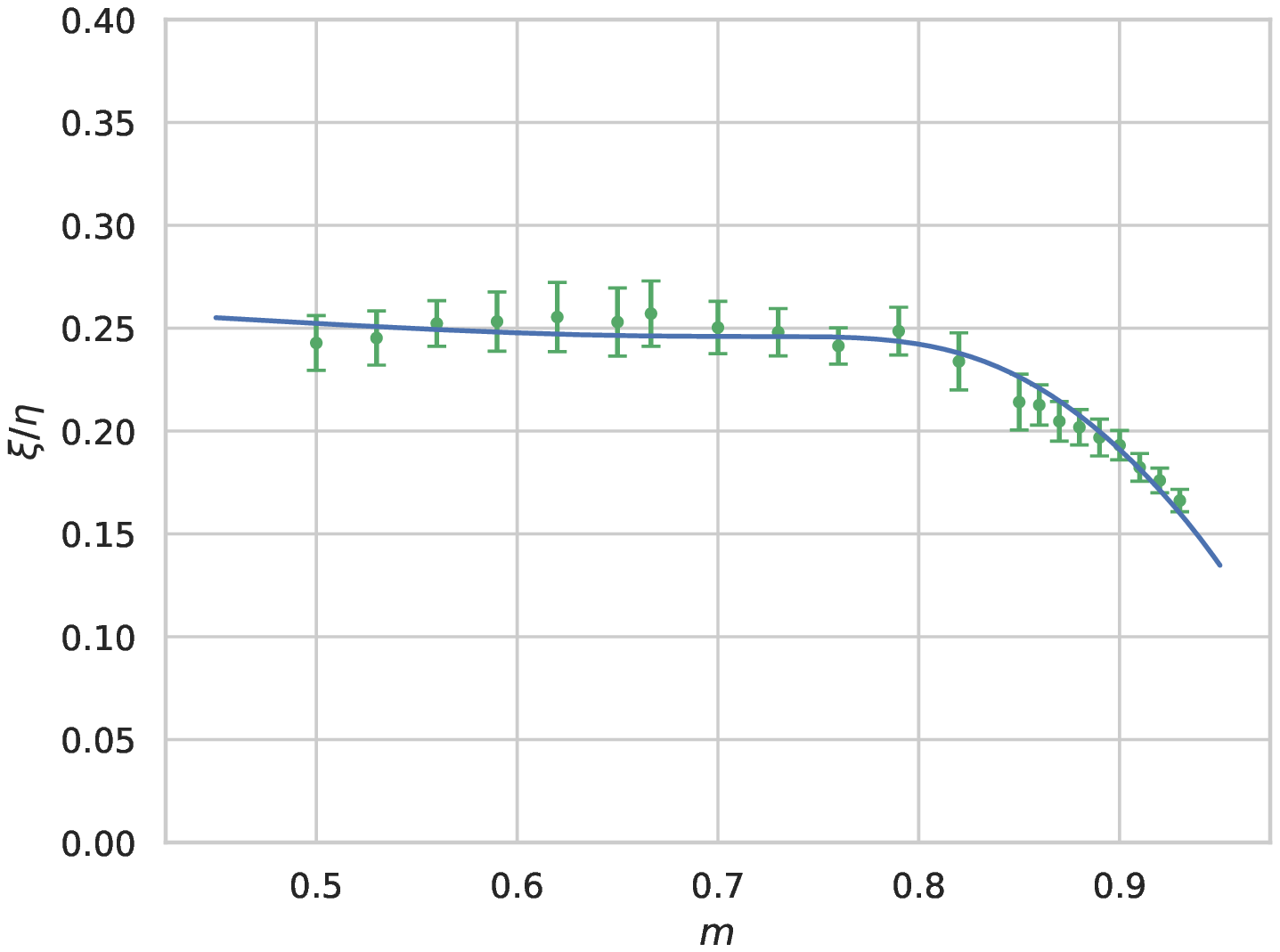}
  \includegraphics[width=0.49\columnwidth]{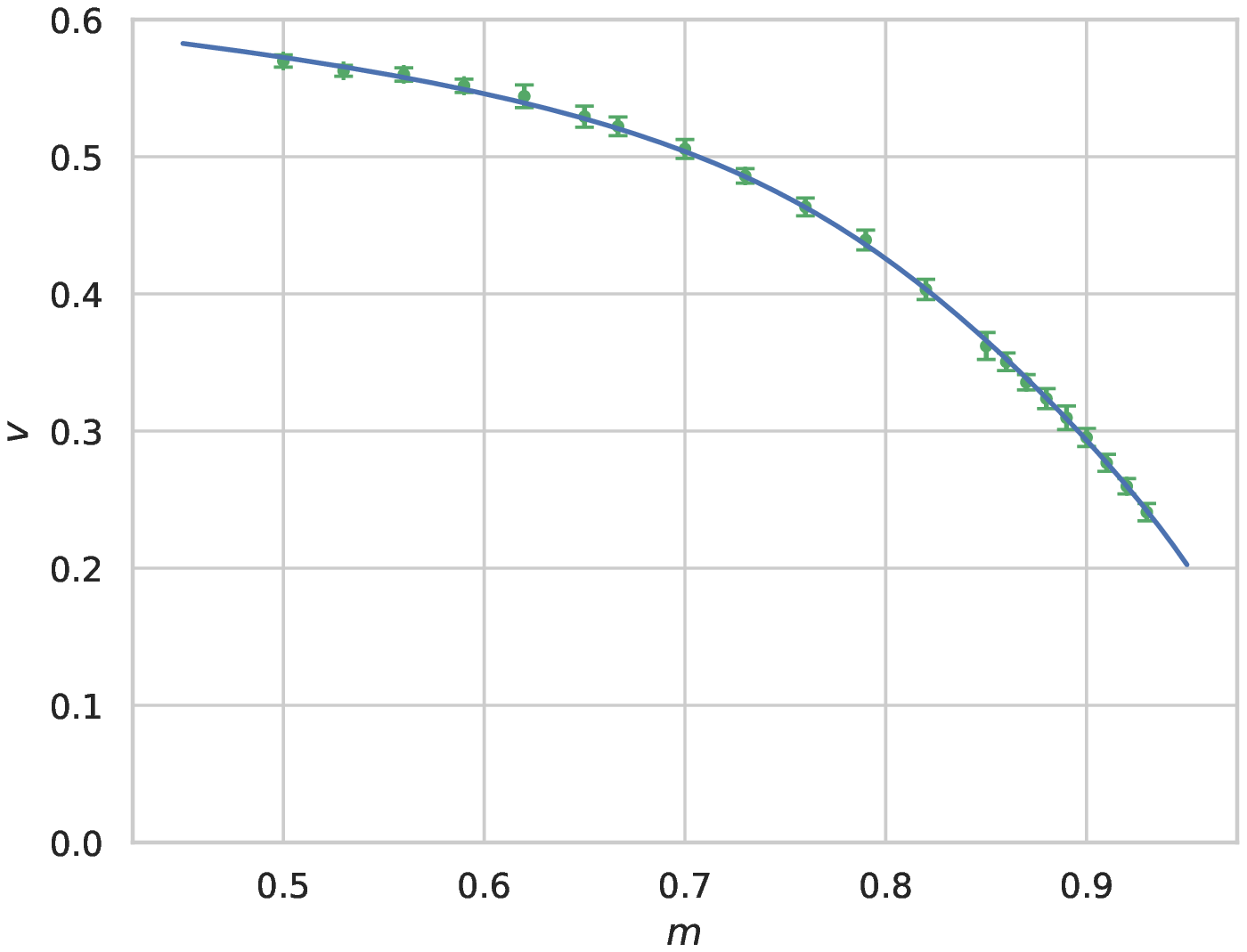}
  \includegraphics[width=0.49\columnwidth]{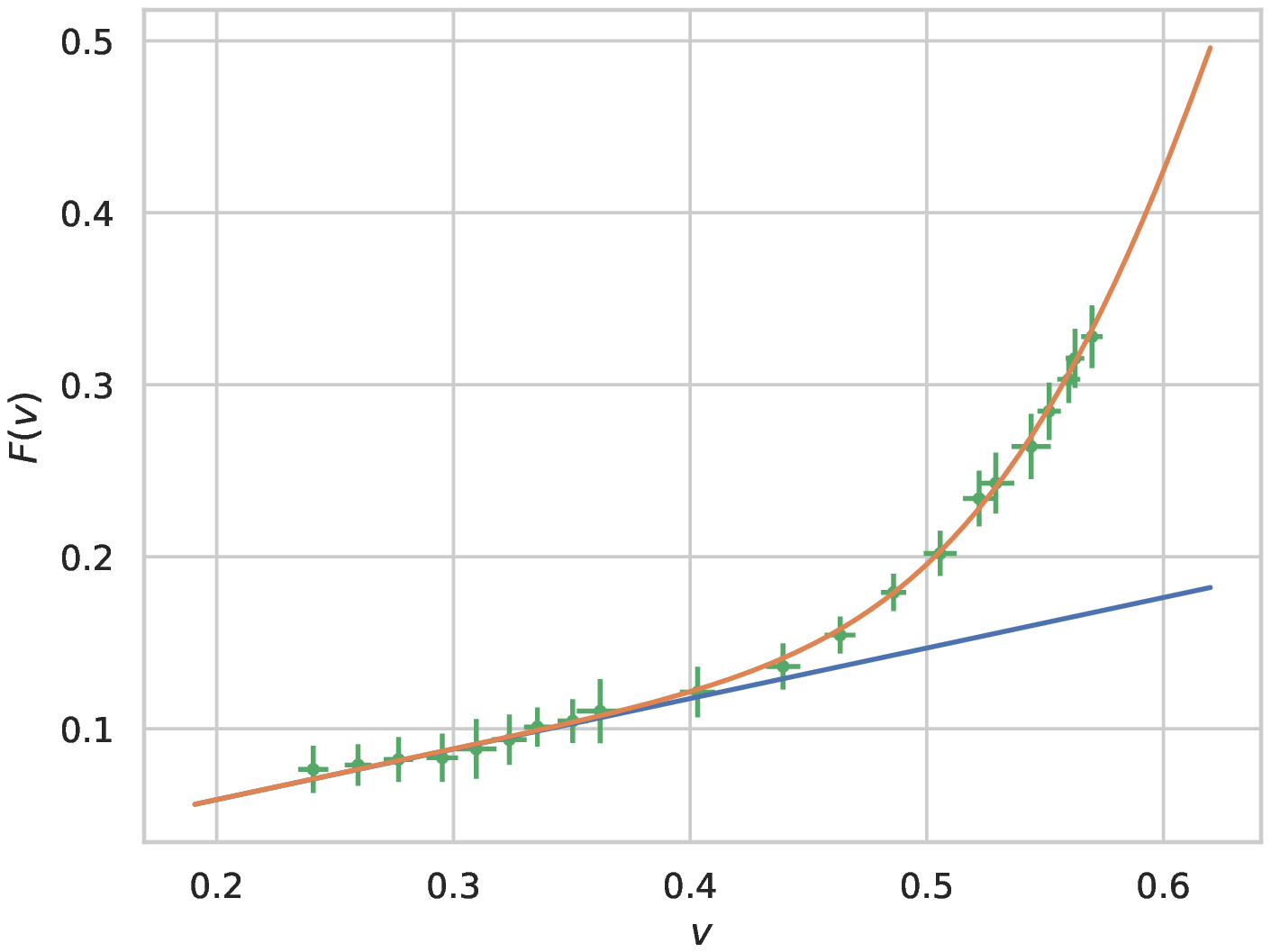}
  \includegraphics[width=0.49\columnwidth]{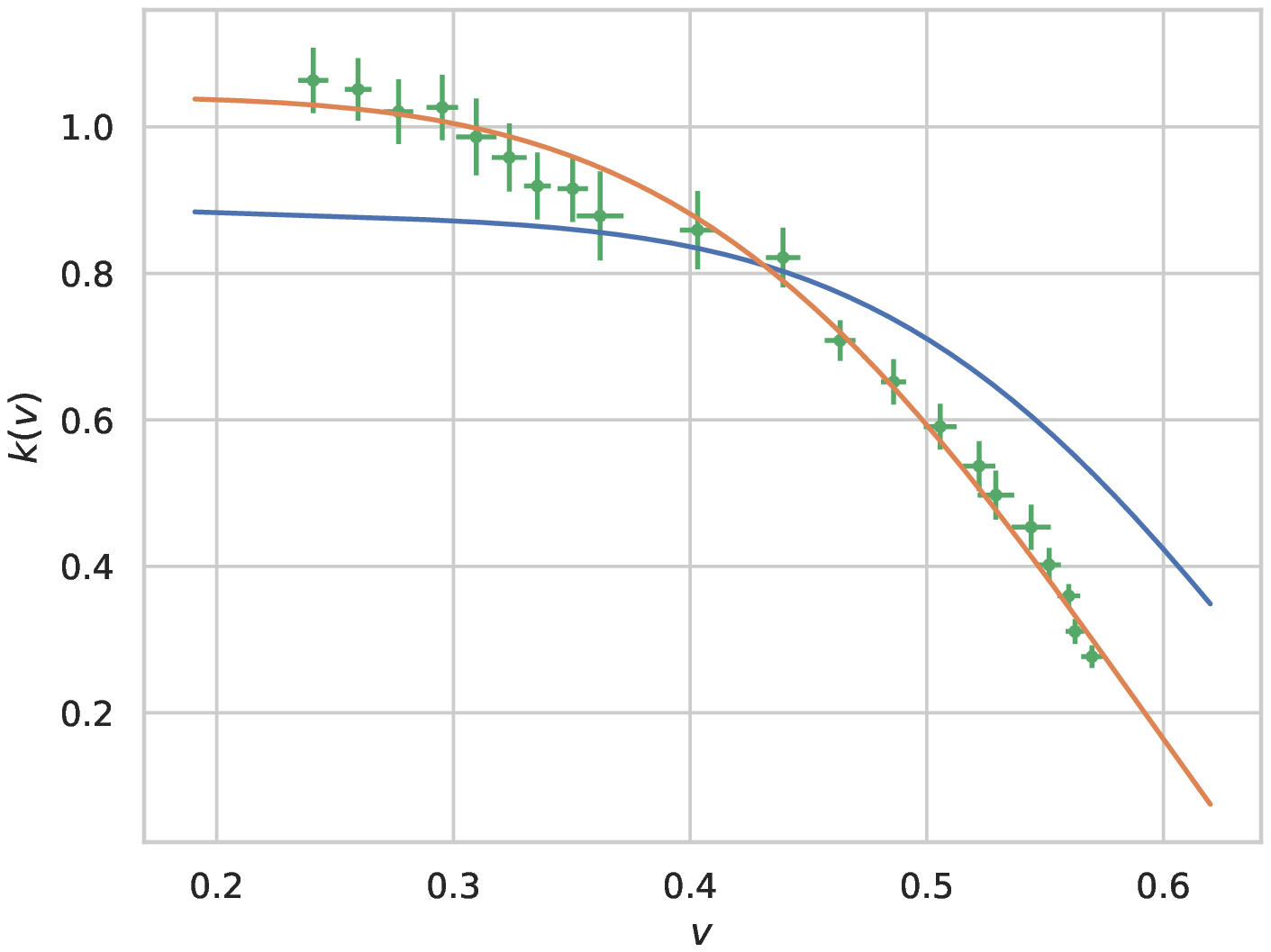}
\end{center}
\caption{A showcase of the VOS predictions for the mean rate of change of correlation length $\xi / \eta$ (top left) and the mean velocity $\langle v \rangle$ (top right) with the solid blue lines corresponding to the calibration and the green points to the measured quantities with statistical $1-\sigma$ uncertainties. The bottom plots show the functions $F(v)$ and $k(v)$ with green points showing measured values and the blue lines showing the standard VOS prediction with the extended VOS prediction in orange.}
\label{figVOSpred}
\end{figure}

The calibrations performed with the scalar field velocity estimator mostly reduce uncertainties (with one exception) as the lattice spacing is reduced, as can be inferred from the lower half of table \ref{table2} and from the contours of figure 8 from Ref. \citenum{Correia:2021tok}. The only exception to this is the parameter $q$ which is again expected, as it is related to the maximal defect velocity and the estimator changes mostly in the high velocity limit. 

Thus lowering lattice spacing brings both calibrations, with different velocity estimators, into better agreement with the only different parameter being $c$. Overall this means one of the most important parameters from an observational point of view, $c$, is the most sensitive in the model, being somewhat affected by resolution effects and choice of velocity estimator.

\begin{table}
\tiny
\tbl{Calibrated VOS model parameters for our two choices of lattice spacing $\Delta x$ and corresponding lattice sizes, for the two different choices of velocity estimators, $\langle v^2_\omega \rangle$ and $\langle v^2_\phi \rangle$, further described in the main text.}
{\begin{tabular}{lcccccccc}
\toprule
\thead{Lattice\\size} & $\Delta x$ & \thead{Velocity\\estimator} & d & r & $\beta$ & $k_0$ & q & c \\
\colrule
$2048^3$ & 0.5 & $\langle v_\omega^2 \rangle$ & $0.37^{+0.02}_{-0.02}$ & $1.27^{+0.17}_{-0.15}$ & $2.33^{+0.21}_{-0.20}$ & $1.21^{+0.03}_{-0.03}$ & $2.57^{+0.06}_{-0.06}$ & $0.03^{+0.02}_{-0.03}$  \\
$4096^3$ & 0.25&                              & $0.34^{+0.07}_{-0.05}$ & $2.32^{+0.52}_{-0.40}$ & $2.62^{+0.29}_{-0.26}$ & $1.06^{+0.03}_{-0.02}$ & $2.37^{+0.06}_{-0.07}$ & $0.25^{+0.02}_{-0.02}$  \\
\colrule
$2048^3$ & 0.5 & $\langle v_\phi^2 \rangle$   & $0.33^{+0.05}_{-0.04}$ & $1.86^{+0.39}_{-0.32}$ & $2.65^{+0.28}_{-0.26}$ & $1.05^{+0.03}_{-0.03}$ & $2.84^{+0.08}_{-0.08}$ & $0.31^{+0.02}_{-0.02}$  \\
$4096^3$ & 0.25&                              & $0.36^{+0.09}_{-0.06}$ & $2.56^{+0.64}_{-0.50}$ & $2.69^{+0.30}_{-0.27}$ & $1.04^{+0.03}_{-0.02}$ & $2.47^{+0.07}_{-0.07}$ & $0.30^{+0.02}_{-0.02}$  \\
\botrule
\end{tabular}}
\begin{tabnote}
Displayed values correspond to 16th, 50th, 84th percentiles of the posterior distributions.
\end{tabnote}\label{table2}
\end{table}

To conclude we present the VOS predictions of the best-case scenario calibration, with lattice $4096^3$, spacing, $\Delta x = 0.25$ and velocity estimator $\langle v_\phi^2 \rangle$ in Fig. \ref{figVOSpred}. The VOS model predicts reasonably well, for all expansion rates, the values of $\xi/\eta$ and $v$. We can see the measured value of $k(v)$ and $F(v)$ seem to be well described by the generalized forms, while the standard forms fail to predict the proper velocity dependency.

\section{Observational Impact}

We will now ahighlight the impact of different calibrations on observational footprints of cosmic string networks. Note that a detailed study is beyond the scope of this manuscript, and this is merely illustrative of the need for an accurate calibration \cite{Correia:2021tok}.

We can use the CMBACT4 code \cite{Pogosian:1999np,Charnock:2016nzm} to compute the Cosmic Microwave Background anisotropies generated by a string network, assuming different VOS calibrations. This code assumes the Unconnected Segment Model \cite{Pogosian:1999np}, where ensembles of straight, randomly oriented segments with separations and velocities given by the VOS model are used to compute power spectra.

We cancompare four different calibrations: standard VOS compatible with Nambu-Goto, standard VOS calibrated for Abelian-Higgs and two extended VOS calibrations present in the previous section, a worst ($1024^3$, $\Delta x=0.5$, $\langle v_\omega^2 \rangle$) and best ($4096^3$, $\Delta x=0.25$, $\langle v_\phi^2 \rangle$) case scenario. The resulting $TT$, $TE$, $EE$ and $BB$ can be found in figure 9 of Ref. \citenum{Correia:2021tok}, all normalized to a string tension $G\mu=1.0\times10^{-7}$. It is clear that the Abelian-Higgs spectra are in better agreement between each other, than the Nambu-Goto one, which is expected \cite{AS,VVO,BB,Blanco,Lazanu1,Lazanu2}.

However, there are still some noticeable differences between Abelian-Higgs calibrations. The most discrepant overall is the $1024^3$ calibration---the worst-case scenario. The $4096^3$ calibration and the standard VOS one are in better agreement, although it should be noted there still exist some scale-dependent differences. To exemplify, at  $l=10$, the relative difference between the two $TT$ spectra is around $16\%$, $30\%$ and $11\%$ for scalar, vector, tensor $C_l$, respectively.

The exact quantifying of how much of these differences are uniquely due to the calibrations or due to the USM requires a more in-depth, detailed study. Nevertheless, this illustrative comparison highlights the need for accurate VOS calibrations.

\section{Conclusion}

In this work we used our recently developed multiGPU Abelian-Higgs code and high-end computational resources of Piz Daint to obtain a robust calibration of the canonical semi-analytical model of string evolution. Using high-resolution simulations from lattice sizes ranging from $1024^3$ to $4096^3$ and various choices of expansion rate. We explored the impact of two different numerical choices on the obtained calibrations. This allowed us to not only reduce statistical uncertainties but also to correct for systematic error sources. Since we tested the sensitivity of model parameters to each of these choices we uncovered that the loop-chopping parameter is highly sensitive to both resolution effects and also to choice of velocity estimator. We showed that a minimum dynamic range of about $512.0$ conformal time units and a minimum lattice spacing of $\Delta x = 0.25$ are necessary to minimize the impacts of these systematic error sources.

We then illustrated one of the main motivations to pursue an accurate, high-resolution calibration of the VOS: observational consequences. We specifically showed that different calibrations can give rise to scale-dependent differences on the computed CMB anisotropy power spectra. Although one cannot (at least without a more in-depth study) exclude that these differences are caused by the various approximations of the Unconnected Segment Model, it is still clear that precise and accurate calibrations are necessary to for both current and future constraints.

\section*{Acknowledgements}

This work was financed by FEDER—Fundo Europeu de Desenvolvimento Regional funds through the COM-PETE 2020—Operational Programme for Competitiveness and Internationalisation (POCI), and by Portuguese funds through FCT - Fundação para a Ciência e a Tecnologia in the framework of the project POCI-01-0145-FEDER-028987 and PTDC/FIS-AST/28987/2017. J. R. C. is supported by an FCT fellowship (SFRH/BD/130445/2017). We gratefully acnknowledge the support of NVIDIA corporation with the donation of the Quadro P5000 GPU used in this research. We acknowledge PRACE for awarding us access to Piz Daint at CSCS, Switzerland, through Preparatory Access proposal 2010PA4610, Project Access proposal 2019204986 and Project Access proposal 2020225448. Technical support from Jean Favre at CSCS is gratefully acknowledged.

\eject

\bibliographystyle{ws-procs961x669}
\bibliography{mg16correia}

\end{document}